\documentclass[conference]{IEEEtran}
\usepackage{cite}
\usepackage[pdftex]{graphicx}
\usepackage[cmex10]{amsmath}
\usepackage{amssymb}

\usepackage{times}
\usepackage{array}
\usepackage[tight,footnotesize]{subfigure}
\usepackage[font=footnotesize]{subfig}
\usepackage{graphicx}
\usepackage{mathtools}
\usepackage{microtype}
\usepackage{amsmath}

\usepackage{alltt}
\usepackage{shadethm}
\newshadetheorem{comment}{Comment}

\def\boxx{{\vcenter{\vbox{\hrule height.3pt
          \hbox{\vrule width.3pt height6pt
          \kern6pt\vrule width.3pt}\hrule height.3pt}}\;}}

\def\impos{{\;\vcenter{\hbox{\rule{5mm}{0.2mm}}} \vcenter{\hbox{\rule{1.5mm}{1.5mm}}} \;}}

\def\mlt{\preceq} 

\def\lrarrow{\leftrightarrow \kern-8pt \rightarrow}

\def\2{\frac{1}{2}}

\def\beq{\begin{eqnarray}}
\def\eeq{\end{eqnarray}}
\def\2{\frac{1}{2}}

\def\lrarrow{\leftrightarrow \kern-8pt \rightarrow}

\def\frightarrow{\rightarrow \kern-11pt /~~}
\def\reducesto{\simeq \kern -3pt >}
\def\intersection{\cap}

\begin{document}
\newcommand{\strust}[1]{\stackrel{\tau:#1}{\longrightarrow}}
\newcommand{\trust}[1]{\stackrel{#1}{{\rm\bf ~Trusts~}}}
\newcommand{\promise}[1]{\xrightarrow{#1}}
\newcommand{\revpromise}[1]{\xleftarrow{#1} }
\newcommand{\assoc}[1]{{\xrightharpoondown{#1}} }
\newcommand{\rassoc}[1]{{\xleftharpoondown{#1}} }
\newcommand{\imposition}[1]{\stackrel{#1}{\impos}}
\newcommand{\scopepromise}[2]{\xrightarrow[#2]{#1}}
\newcommand{\handshake}[1]{\xleftrightarrow{#1} \kern-8pt \xrightarrow{} }
\newcommand{\cpromise}[1]{\stackrel{#1}{\frightarrow}}
\newcommand{\policy}{\stackrel{P}{\equiv}}
\newcommand{\field}[1]{\mathbf{#1}}
\newcommand{\bundle}[1]{\stackrel{#1}{\Longrightarrow}}

\title{On The Scale Dependence and Spacetime Dimension of the Internet with Causal Sets}

\author{Mark Burgess\\ChiTek-i AS}
\maketitle
\IEEEpeerreviewmaketitle

\renewcommand{\arraystretch}{1.2}

\begin{abstract}
  A statistical measure of dimension is used to compute the effective
  average space dimension for the Internet and other graphs, based on
  typed edges (links) from an ensemble of starting points.
  The method is applied to CAIDA's ITDK data for the Internet. The effective
  dimension at different scales is calibrated to the conventional
  Euclidean dimension using low dimensional hypercubes. Internet
  spacetime has a `foamy' multiscale containment hierarchy, with
  interleaving semantic types.  There is an emergent scale for
  approximate long range order in the device node spectrum, but this
  is not evident at the AS level, where there is finite distance
  containment. Statistical dimension is thus a locally varying
  measure, which is scale-dependent, giving an visual analogy for the
  hidden scale-dependent dimensions of Kaluza-Klein theories. The
  characteristic exterior dimensions of the Internet lie between $1.66
  \pm 0.00$ and $6.12 \pm 0.00$, and maximal interior dimensions rise
  to $7.7$.
\end{abstract}

\hyphenation{before}
\hyphenation{immun-ology}


\section{Introduction} 

The Internet is large graph for which we can explicitly and
empirically measure the average statistical space(time) dimension.
The concept of a statistical dimension was introduced in
\cite{myrheim1}, and has since attracted interest in causal set theory
and discrete approaches to quantum gravity
\cite{myrheim1,meyerthesis,causeset1,causeset2}. There is dual scientific
interest in determining the dimension of an actual phenomenological
network: it helps to set aside any prejudices about what dimension we
might expect from a virtually interacting processes by straightforward
measurement, and it allows us to calibrate the technique for a more
intuitive assessment using graph theoretic methods. With the
introduction of modern graph databases, this task becomes quite easy.
On a technological level, the average dimension of the Internet indicates
the availability of redundant paths through the mesh and is thus a measure of
the robustness of the Internet to failure.

\begin{figure}[ht]
\begin{center}
\includegraphics[width=8.5cm]{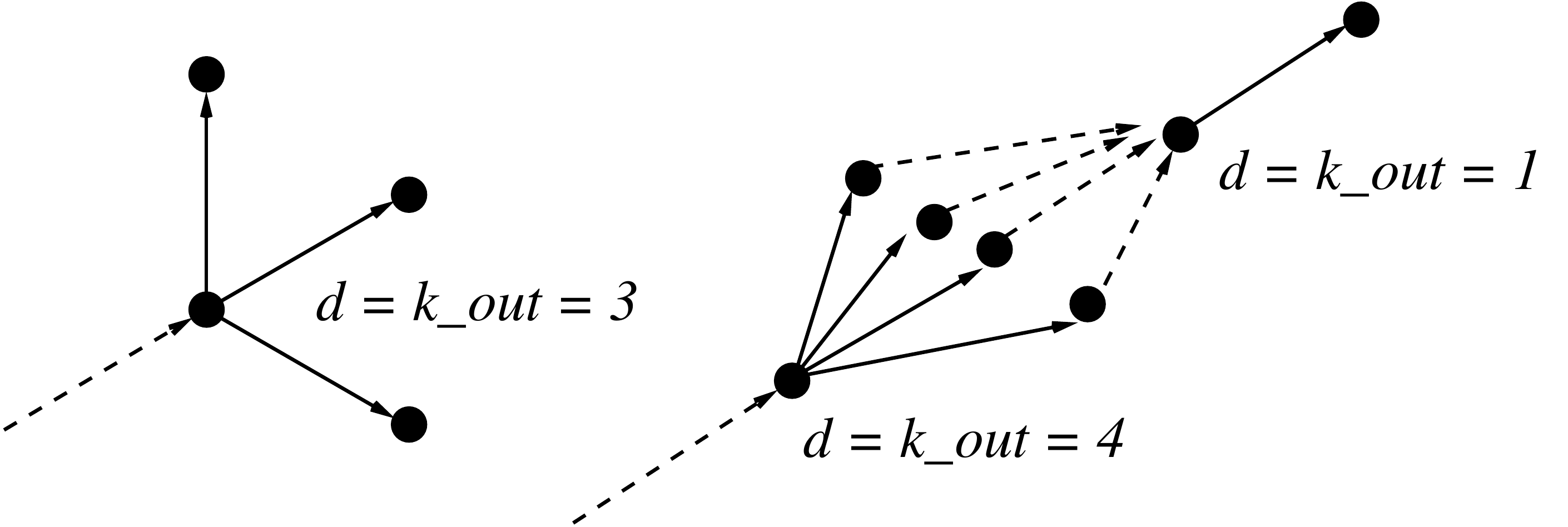}
\caption{\small Graph dimension. The number of degrees of freedom available
as paths may vary throughout a graph. The number of alternatives is also scale dependent.
Over short distances, different paths may be irreducible, but over longer times and distances
they may converge back together (like a multi-slit experiment).\label{dimension}}
\end{center}
\end{figure}

Conventionally, the idea of space(time) dimension is rooted in long
standing traditions of Euclidean geometry and Cartesian vector spaces.
This is a geometric idealization, inferred more by the convenience of
symmetry than observed directly. In physics, the underlying structure
of spacetime is reasoned to be discrete on some scale (the Planck
scale is typically assumed), but the dimension of a discrete spacetime
is not as simple to define as in a continuous Euclidean space.  We
also have to deal with different `semantics', or the roles for which
we count dimensions. For example, the `kinetic dimension' of a space
is conventionally equal to half the number of independent spatial
degrees of freedom (or forward directions) available to a process (see
figure \ref{dimension}). Here, there is an implicit notion of
continuity of momentum in selecting forward directions only.  In solid
state and lattice field theories of physics, a finite spacing is
maintained between points in the parameter space to avoid infinities
associated with the differential coincidence limit, and to separate
the effects of bound states.  This makes a direct connection between
classically Newtonian causality and computational processes over
multiple timescales.

Definitions of dimension vary, but most stay close to a tradition in
philosophy of studying trajectories of rigid bodies within an
idealized ambient `theatre'. The parameters characterizing such paths
are continuous real-valued coordinates.  The success of this approach
was affirmed by Newton's treatise on mechanics and many later
theories, which have etched the truth of this model into the
collective consciousness.  However, the attractively simple idea of a
Euclidean vector space is insufficient to characterize the effective
degrees of freedom of more general processes, including those found in
biology and computer science, where the parameters form discrete
graphs, so its a happy occasion to find phenomenological data to
study.  In graph theory, the dimension of a graph may be defined in
several ways: both in terms of the same dynamical counting of degrees
of freedom above, and even by the embedding of the graph into a
Euclidean space as a last resort.  The latter is popular in developing
approximations in network Machine Learning for instance, because it
keeps reasoning within a familiar Pythagorean regime.

\begin{figure}[ht]
\begin{center}
\includegraphics[width=7cm]{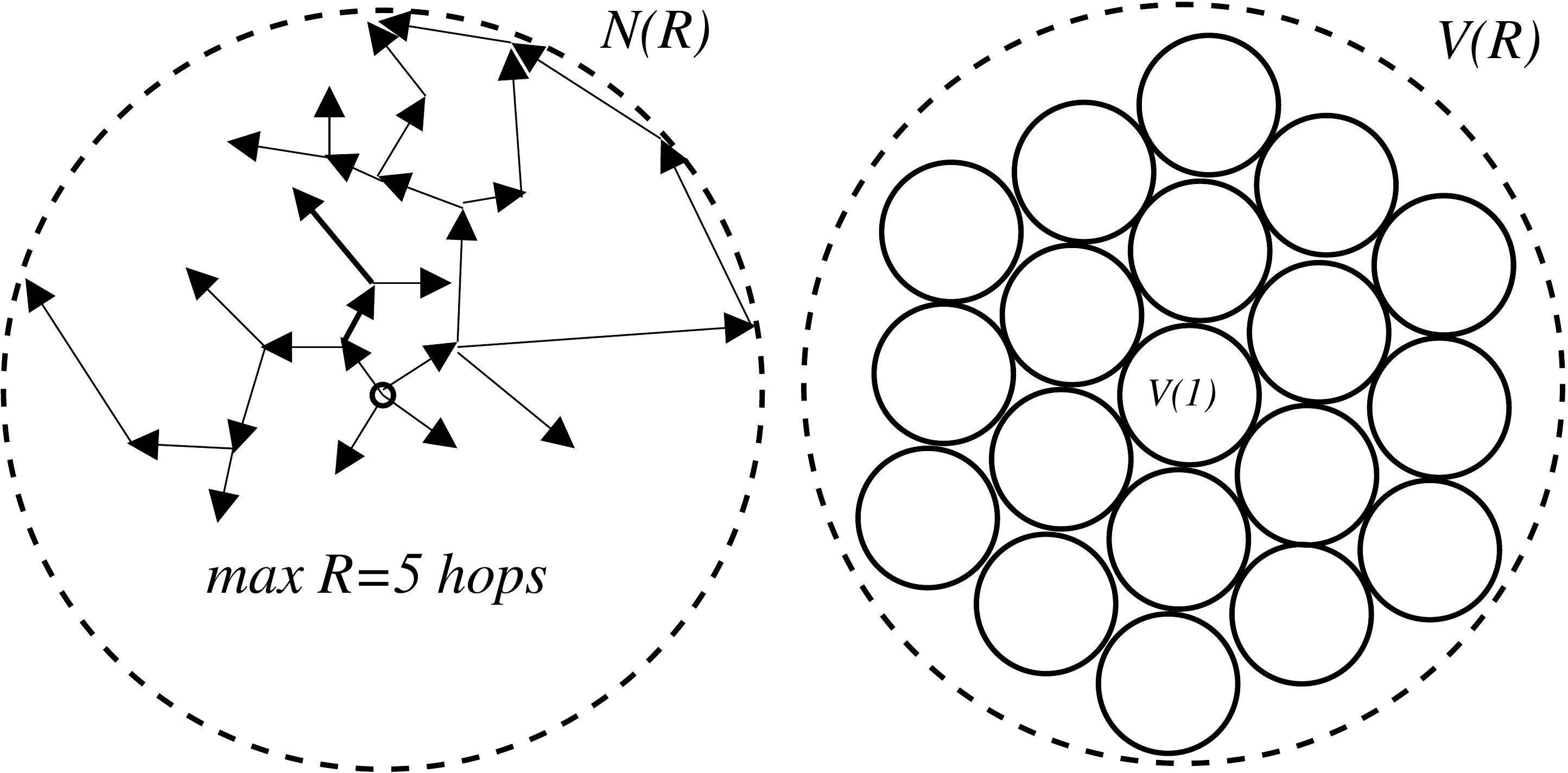}
\caption{\small Graph statistical dimension over an imagined average
  uniform volume by counting nodes reachable within $R$ hops. The role
  of direction is no longer relevant in a radial probing of a ball
  about the origin.\label{dimension2}}
\end{center}
\end{figure}

\section{Measuring graph space}

In a directed graph, locations are nodes (vertices), and the forward or
outgoing degree $k_\text{out}$, or number of links (edges) are the dimension
at that point.  This idea is scale dependent (figure
\ref{dimension} again), which may feel unfamiliar--though a similar idea is invoked
in modern Kaluza-Klein theories, where hidden degrees of freedom are
considered to be short range or `rolled up' on a small scale.

Most lay-persons probably think of the Internet as a network of cables
stretched around the globe, but this is not a representation of its
dynamics.  The functional behaviour of the Internet is built on the
outcomes of interfering virtual phenomena, just as the behaviour of a
quantum multi-slit experiment is based on the behaviour of waves not
the slits by themselves.  Empirically, we probe the Internet as a
graph, from within the graph itself, by sampling data about
neighbouring nodes from within a `radius' defined as a certain number
of hops along graph edges from a given random origin.  The collection
of the data is non-trivial, and involves assimilating data from
multiple patches into a covering; but, luckily the data exist at the
Center for Applied Internet Data Analysis (CAIDA)\cite{midar}. By
letting a known process run over all possible acyclic trajectories
from the origin, and looking at reflections---a bit like a sonar map.
In the case of the Internet, one uses the ECHO protocol, or the
well-known traceroute program.  These reflections are then collated
and averaged to create a map, which is held in a graph database
(ArangoDB). The resulting map is somewhat fictitious, being a
representation of a sum over multiple ephemeral dynamical
processes--in much the way that a television cathode ray image is a
convenient illusion, or a quantum wavefunction is a representation of
possible energy channels, because its points are never observed
simultaneously--we rely on a persistence timescale to collate and
define effective simultaneity.

To probe the dimension of the map, we can now repeat the same process on the
aggregate map, using the database to search for all nodes reachable by
acyclic paths within a certain number of hops.  These data samples are
deterministic, and thus without uncertainty, but they are not local
and representative of the whole graph.  They can be collected into
ensembles by randomizing the origin---choosing sample sets which have
either closely separated origins or widely separated origins to study
the locality of characters. In each case, we can easily retrieve the
number of reachable points $N(R)$ as a function of integer number of
hops $R$ with a single database query. This can now be reported on
average, with associated error bars.  The effect is to probe spacetime
from the perspective of `open balls', mimicking the basic topological construction of a
manifold, albeit with discrete integral adjacent jumps.  In essence,
we deal only with time-average point correlations, and the
directionality of the original probes is averaged away by the
methodology. This may be presumptuous, if we're interested in
behaviour on a detailed level, especially when inhomogeneous
characteristics of processes are taken into account, but it fits most
people's intuitions of what a space should be\cite{spacetime1}.

\begin{figure}[ht]
\begin{center}
\includegraphics[width=9cm]{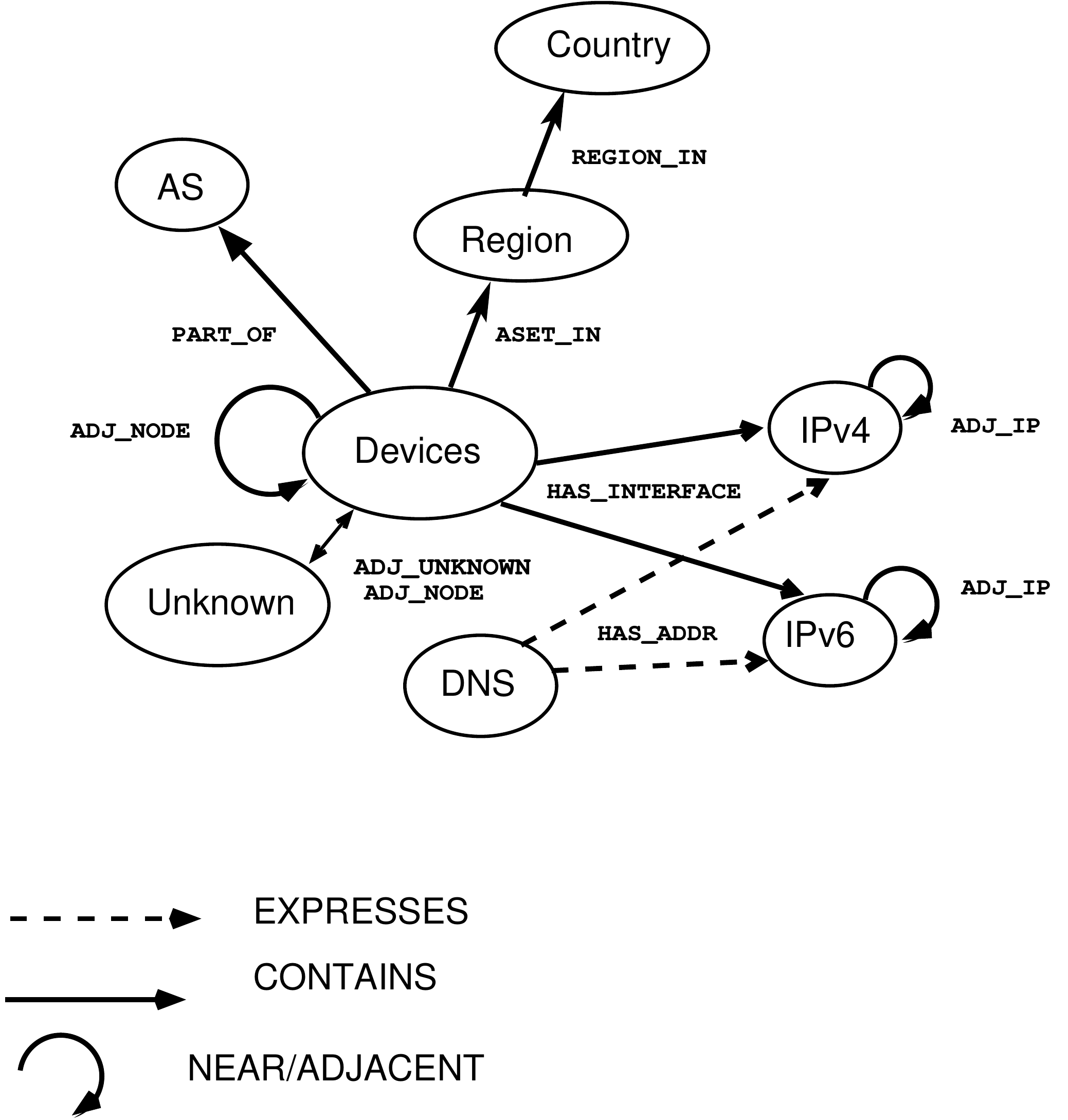}
\caption{\small Semantic spacetime type model for ITDK data. A model
  of ordinary spacetime would consist of only a single type set with a
  single loop arrow. The ITDK data represent spacelike hypersamples over multiple types of node,
  so there are no timelike connections. The looping arrows represent
  `Near' associations, which are assumed directionless. The straight
  arrows are semantic containments and property expressions, which
  contain no topological information, but allow us to infer
  dimensional connectivity between larger coarse grained location
  concepts.\label{itdk_model}}
\end{center}
\end{figure}

This process assumes that all reachable nodes are of homogeneous type,
and that all links are traversable in both directions. Both these
assumptions are presumptuous in the case of the a general network.
In collating the data, we have applied a semantic spacetime model to
separate out independent semantic degrees of freedom which allows us to probe
the graph through different virtual process channels (see figure \ref{itdk_model}).
These channels can be associated with different protocols (which are the
computer science equivalent of force carriers in physics). Different types of
connection and different kinds of node have different semantics, and
we can (indeed should) separate them. However, there is also an
effective picture of what it means for nodes to be adjacent.  

In what follows, we look at two channels: one that correspond
approximately to network `devices' (what CAIDA call Nodes, or entities
that appear to share IP interfaces) and so-called Autonomous Systems
(AS) which represent large organizations hosting the devices.  Owing
to the unobservability of certain regions, there are also `wormholes'
or `hyperlinks' where traces may appear to tunnel through several
regions without being observable.  We take those into account, but
they distort the underlying picture with an additional virtual
phenomenon. The channels communicate by the IP protocol and the BGP
protocols respectively. BGP travels virtually on top of IP, but is
still semantically independent of it. We can perform the same measurement process on each
independent graph to measure a result.

\begin{figure}[ht]
\begin{center}
\includegraphics[width=7cm]{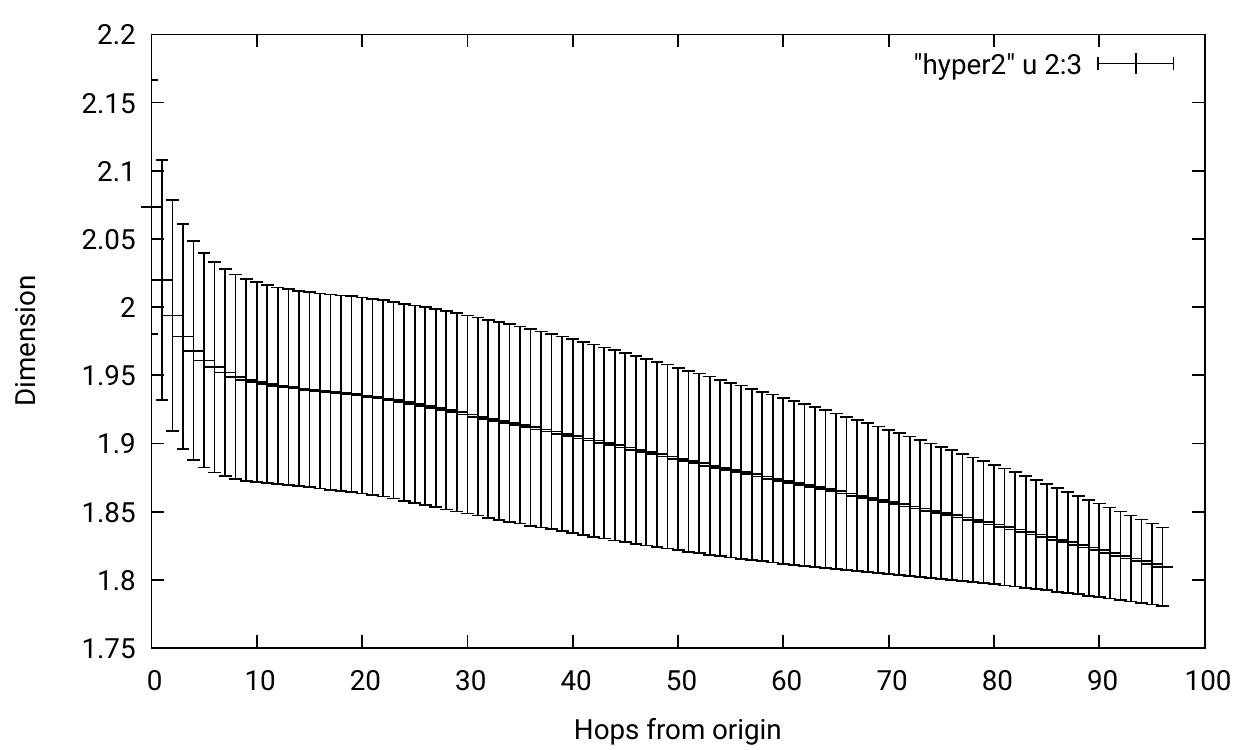}
\caption{\small 2-cube dimension as function of hops from a random
  origin, using the random sampling method for
  calibration.\label{hyper2}}
\end{center}
\end{figure}

\begin{figure}[ht]
\begin{center}
\includegraphics[width=7cm]{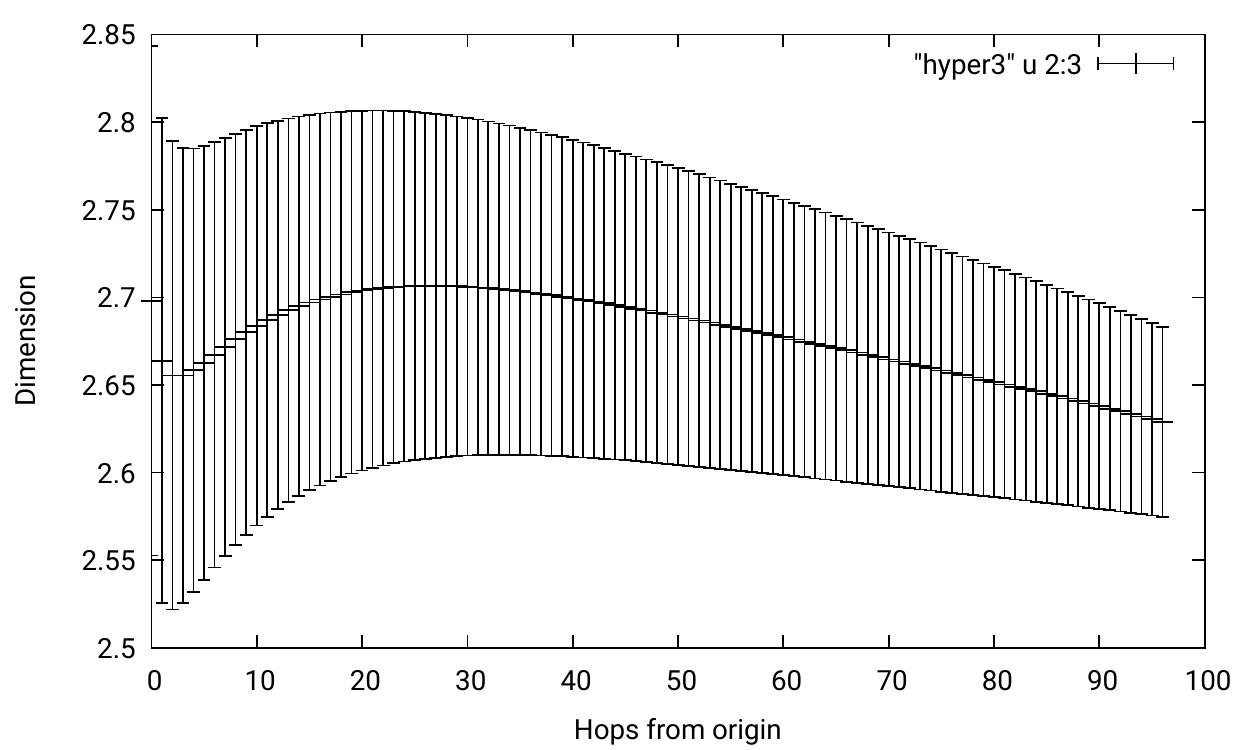}
\caption{\small 3-cube dimension as function of hops from a random origin, using the random sampling method for calibration.\label{hyper3}}
\end{center}
\end{figure}

For calibration of a statistical method, we remark here that regular
(crystalline) lattices (figure \ref{cubes}), exhibiting long range
order, have a special significance for our intuition about spacetime
dimension, since these are ubiquitous and for the basis for Cartesian
coordinates, thus we use these for calibrating our measure. However,
the likeness of a graph to a lattice or other continuum manifold is of
mainly heuristic interest in the case of more complex multiscale
systems, like the Internet, since the concept of dimensionality is
mainly associated with predicting degrees of available states and
dynamical freedoms locally.  Nevertheless, it is of interest to test
the method and the intuitiveness of its results.  The study closest to
our, in spirit, is perhaps the use of causal sets to explore the
science citation process \cite{csdimension2016}.

\section{Related physics}

A brief contextual placement for measuring dimension by this method is
in order.  Studies of graph-like spacetimes in physics nearly always
look for the emergence of manifold (quasi-Euclidean) behaviour from
random probabilistic process distributions, since their purpose is to
imagine the emergence of our quasi-Euclidean large scale experience in
some limit. This is not our purpose here.  Directed graphs are more
like causal (kinetic) processes of Feynman diagrams or other process
diagrams. These are of direct interest to Internet design, as well
being suitable for comparison with other dynamical systems.

Continuum space is a precise abstraction, but discrete spacetime is
less constrained by requirements of symmetry, so we are cautious in
making assumptions.  Myrheim \cite{myrheim1} was characteristically
early in pointing out that for discrete spacetime we can only define
dimension statistically.  Detailed mathematical work by Meyer
\cite{meyerthesis} developed a language for so-called Causal Sets as
an abstraction based on a theoretical model for node adjacency; this
used Poisson distributed sprinklings of nodes to form directed graphs.
Although arbitrary, this allows results to be calculated analytically.  A more
recent mathematical formulation in \cite{webofchains2019} exploring
the probabilistic aspect of point filling and its relationship to
manifold structures.  In these modern renditions, one refers to
Alexandrov sets over the limits of a causal interval $\cal C$.  This
is defined over an interval of values $p,q$: $I(p,q) \equiv \{r | p
\mlt r \mlt q\}$, for finite cardinality.

In our case, the data form a cumulative map of such channels, which is
a summary of many overlapping probes, in which the boundary conditions
may be both quickly or slowly varying.  For this to make sense, we
infer the notion of persistent channels of communication (edges)
between the quasi-pointlike vertices of the graph.  We wish to
preserve these explicit communications channels in the model, with an
effective connectivity given by the overlap of their promised open
sets: $\pi^+ \intersection \pi^-$, which is equivalent to the
Alexandrov set $A(x,y)=I^+(x)\intersection I^-(y)$ in the
continuum\cite{promisebook,webofchains2019}. This leads to rather
different topologies than the scale-free Poisson processes, and requires some
modification of the approach to estimating bulk space-filling.

In nearly all studies of statistical dimension, the points, which are
effectively nodes or vertices of a graph, are all of a single
featureless kind---formally `typeless' in a computer science sense and
are generated probabilistically. However, this is not the case in
technology, nor in chemistry and biology, etc, where the concept of a
variegated {\em semantic space} becomes useful\cite{spacetime1}.
Spatiotemporal homogeneity is convenient for invoking an implicit
ensemble, i.e.  an accumulation of processes into a distribution that
leads to continuous fractional scaling. These more complex systems are
of interest both on a practical and theoretical level, since they are
much more accessible and can be observed directly, and have very
predictable functional characteristics.  Semantic spacetime
is a (promise) graph theoretical model of discrete processes, with
arbitrary set-valued overlap criteria that may have any number of
types, equivalences, and orderings. Spacelike features are introduced
through the semantics of spatial relationships (adjacency,
containment, interior properties, etc) rather than by quantitative
measure arguments.  The relationship between Semantic Spacetime and
Causal Sets has been discussed in \cite{paper4}.

Our ability to detect dimension depends on the `channel depth'
in the process hierarchy to which we can probe.  Pathways through
space may be resolved from `routes' between locations, captured and
accumulated from virtual and ephemeral transits with long-term
invariant characteristics. The relationship between routes and cables
is not necessarily single valued nor deterministic, just as the
relationship between quantum measurements and spatiotemporal boundary
conditions is not deterministic and may not plumb the true depth
of a process hierarchy. In this sense, an Internet packet
route is quite like a quantum multi-slit experiment, with several
parallel processes interfering over sampling subtimes.  Invariant
links are discerned by the observation of these interior sampling
process channels over the elapsed (exterior) time of the experiment.
This sampling takes many days. The resulting interference paths may be
sewn together into a persistent hypersurface.

\section{The degree distribution}

The first characterization for any graph $\Gamma(\nu,\lambda)$ is its
description in terms of nodes $\nu$ (also called vertices) and links
$\lambda$ (also called edges). Since every graph maps to a process and
vice versa, this is also an initial indication of the dimensionality of
kinetic behaviour.

\begin{figure}[ht]
\begin{center}
\includegraphics[width=7.5cm]{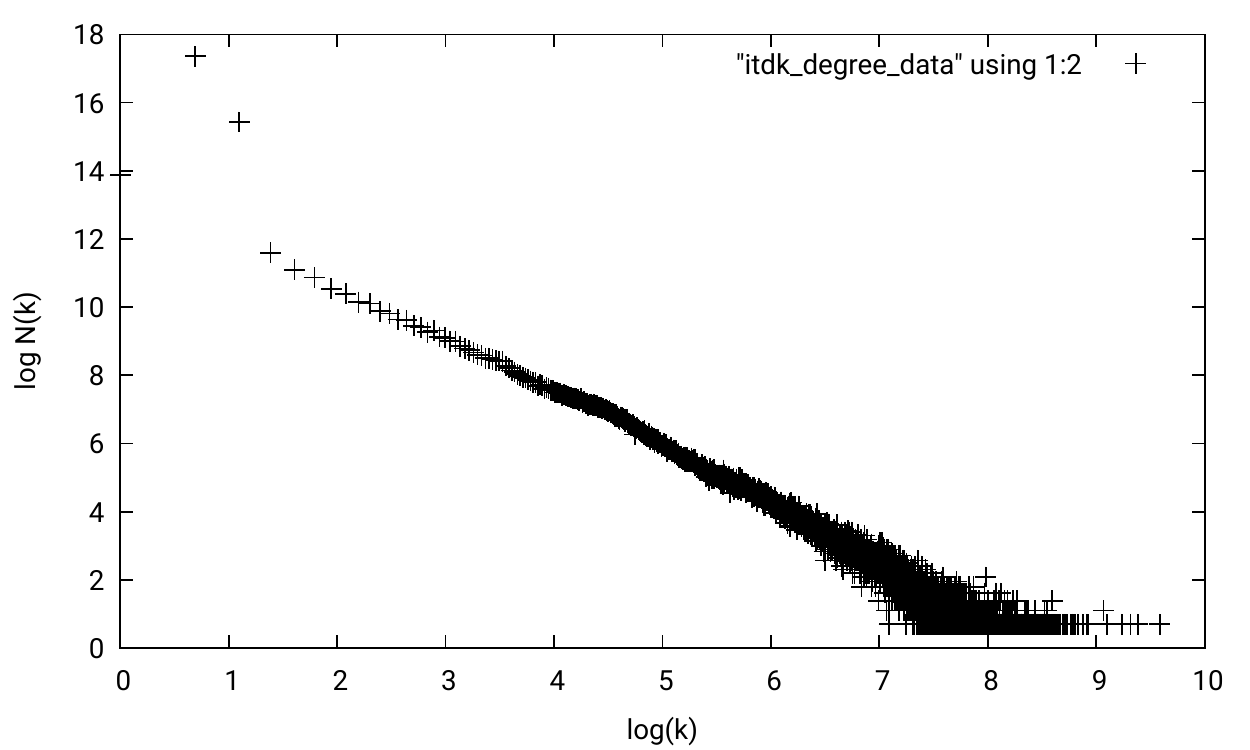}
\caption{\small A log-log plot of the undirected degree distribution
  $(\log(k),\log(N(k)))$ for the Internet Device node
  channel, with $1 \le k \le 8658$.\label{itdk_degree}}
\end{center}
\end{figure}

In a continuum vector space, the number of independent local basis
vectors yields the usual dimension of space. In a continuum, these
basis vectors are only completely independent when they are orthogonal
to one another. The continuity (or relative absence of holes) allows
us to make sense of dimension at a single point, as a limit. The
homogeneity of the continuum allows us to characterize the dimension
for the whole space, as the same everywhere. Graphs, however, are different at every point.
The closest corresponding idea in a graph is the {\em node degree} $k$
at a point. The independence of neighbouring nodes is obvious at one
hop, but their independence over many hops remains to bee seen. How do
we know that two directions are not mainly parallel over long
distances and arrive at the same final destination? We can't know this
except statistically; counting exact trajectories and looking at their
interference patterns is possible in principle, but impractical in
practice for a discrete spacetime.

For a directed process, the outgoing node degree $k_\text{out}$ is the
number of edges leaving the node, following the implicit ordering of
the directed edges. This is usually what we want to use for
trajectories.  Any class of edges constitutes a partial ordering of
the nodes they connect, and the sum of outgoing possibility leads to a
`future cone' of reachable points.  However, in a graph (at each point) different outgoing
paths are always initially independent, as long as they lead to different
neighbouring nodes, because there is no way to deform one edge into a
sum of others without passing through some kind of speculative, i.e.
non-existent, embedding space in which we visualize it. The ability to
do so may depend on the number of links over which we try (see figure
\ref{dimension}). The number of incoming and outgoing edges that
connect a node (vertex) to its neighbours can be different at every
single node.  Thus, the effective {\em local dimension} at a point can
be different at every location too.

\begin{figure}[ht]
\begin{center}
\includegraphics[width=5cm]{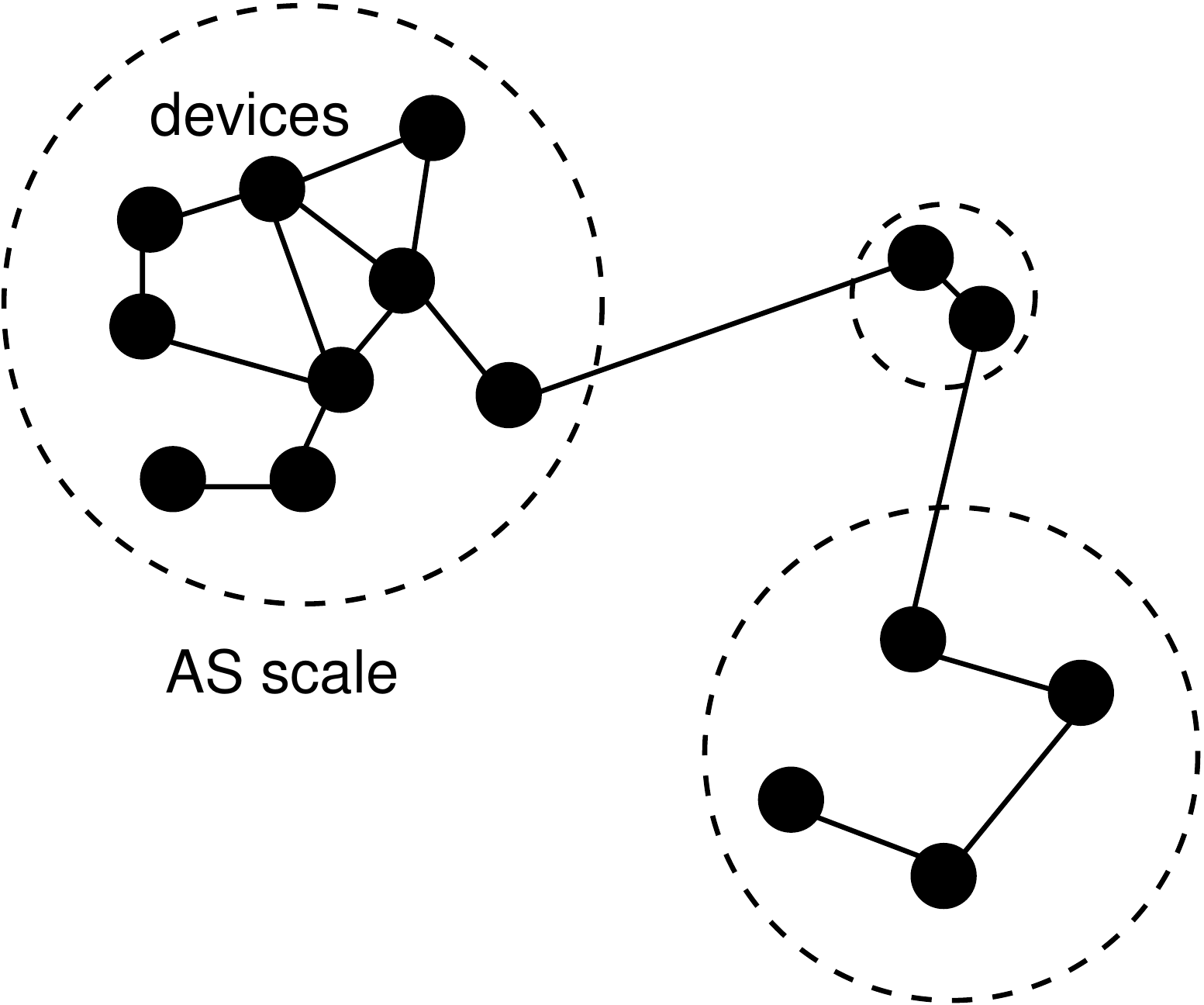}
\caption{\small The hierarchy of containment in the Internet
  structures.  Devices are the smallest connective locations with
  point to point $IP$ connections.  (AS) Autonomous Systems are formed
  from supernode collectives of Device nodes. Independent connections between
  these are maintained through BGP.\label{devAS}}
\end{center}
\end{figure}

Counting the half average total node degree distribution $N(k)$ for the Devices channel and computing
the dimension at one hop.
\beq
2\times D(1) \equiv \langle k\rangle = \sum_i  \; k_i p(k_i),
\eeq
where $p(k_i) = N(k_i)/\sum_i N(k_i)$.
For mutual information channels, we can simply take half $\langle k\rangle/2$
This gives an average local point dimension of $\langle k\rangle/2 = 1.655884$
for the device channel.
This definition of dimension is the relevant microscopic dimension for
instantaneous stepwise dynamics, but it isn't representative of large
scale phenomena.  On a larger scale, path degrees of freedom that
appear independent over shorter path segments might not be independent
over longer segments (figure \ref{dimension}).  Consider a path that
forks into two but then reconverges into a single path after some
distance. In this case, the dimension at point $A$ is two, but the
dimension on a larger scale is only 1, since all paths lead to a
single direction.
We can measure the characteristics of the graph on different semantic
as well as dynamical scales (see figure \ref{devAS}), from its hierarchical
composition.

\begin{figure}[ht]
\begin{center}
\includegraphics[width=6cm]{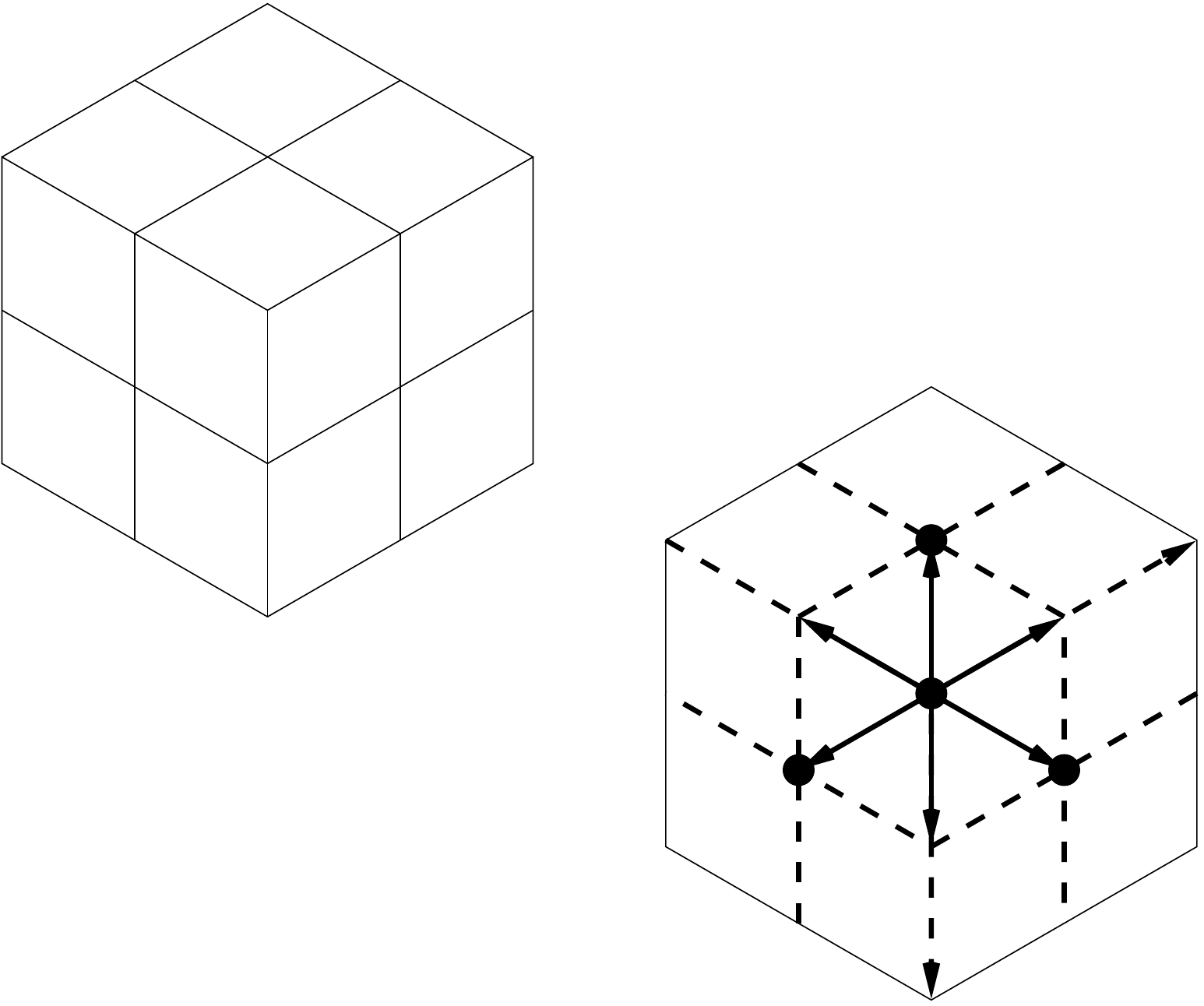}
\caption{\small Graph geometry and embedding. The theoretical notion of a cube of radius
$L$ (left) is different from the empirical reach of a spanning tree (right).\label{cubes}}
\end{center}
\end{figure}

\begin{figure}[ht]
\begin{center}
\includegraphics[width=7.5cm]{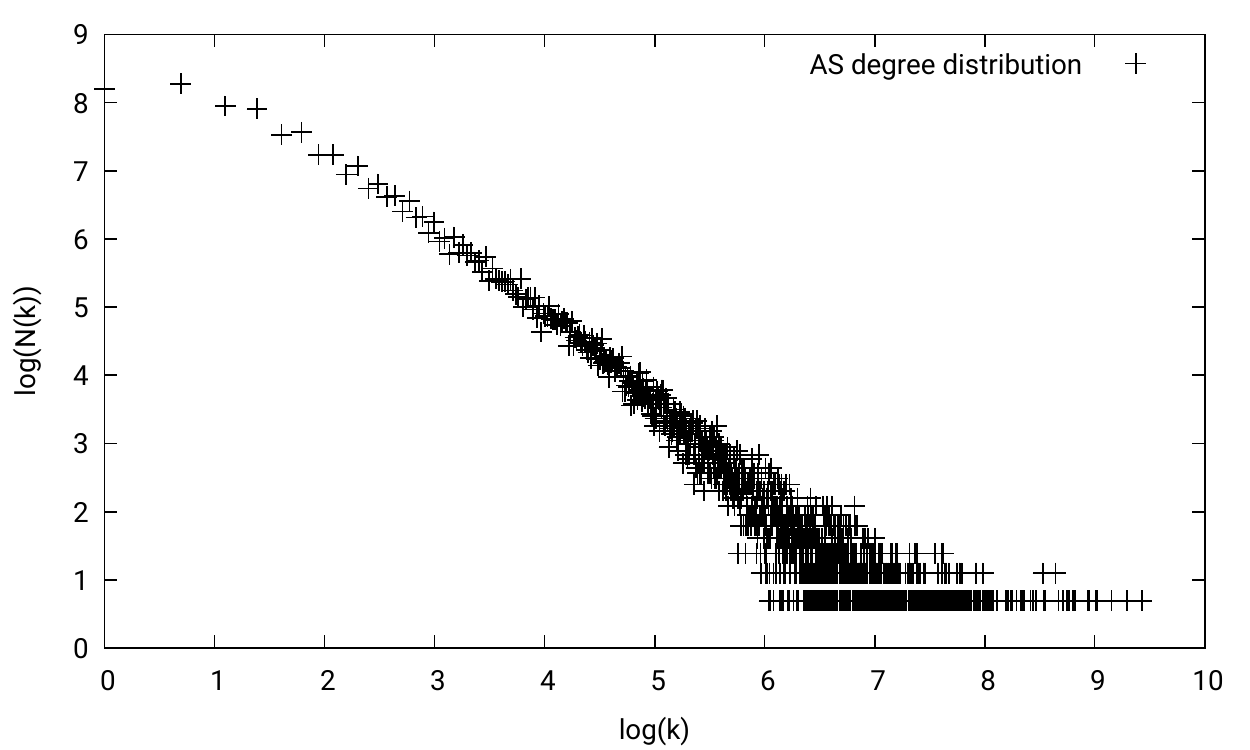}
\caption{\small A log-log plot of the undirected degree distribution
  $(\log(k),\log(N(k)))$ for the Internet AS supernode
  channel, with $2 \le k \le 7621$.\label{itdk_degree_AS}}
\end{center}
\end{figure}

The degree distribution for AS supernodes (containing each many device
nodes) is given in figure \ref{itdk_degree_AS}. For the AS (BGP)
supernodes, the effective dimension as average $\langle k_\text{out}
\rangle/2 = 51.67$. This is significantly higher, since there are many
more outgoing paths between AS providers.

\section{Statistical geometry}

To overcome the scale dependence of node degree, we can look for a
method that can be adapted to any scale with a suitable renormalization.
A statistical approach was proposed by
Myrheim\cite{myrheim1} and by Meyer\cite{meyerthesis}, based on the
assumption of a probabilistic Poisson scattering process.  The
advantage of empirical data is that we don't have to postulate the
existence of a prior probability concept to explore the dynamics of
the Internet's process evolution; however, this means we need a
somewhat different formula for dimension--one that has the semantics
appropriate to small number processes, but which matches our
intuitions about dimension from Euclidean schooling.  A simple form of
this can be understood by imagining space to be an virtual representation
of degrees of freedom, as in kinetic processes.

The effective volume of an $n$-cube of side length $L$ is
just $L^n$, and the volume of a ball of radius $R$ in $n$ dimensions
is:
\beq
V_n = \frac{\pi^{\pi/2}}{\Gamma\left(\frac{n}{2}+1\right)}\; R^n.
\eeq
If we imagine trying to fill a volume of space with unit cubes or
balls, then the number is determined by the dimension of the space (or
vice versa) and some assumptions about the symmetry of volumes. 
We can estimate the effective dimension
of a process by thinking of the vertices of a graph as $n$-balls that
touch when linked by edges. In order for a space to accommodate such
linkage, it requires a minimum number of dimensions, given roughly by
the ratio of volumes. After a certain number of such edges (hops or
links), we can think of the process as a radius from the origin.
The number of nodes after a certain radius us thus:
\beq
N_\text{nodes}(R) = \frac{V(R)}{V(1)} = \frac{R^n}{R_0^n} = N_\text{hops}^n.
\eeq
or
\beq
n \simeq \frac{\log N_\text{nodes}}{\log N_\text{hops}}\label{dimdef}
\eeq
This ratio of logs means that we can only usefully measure the
dimension for hop lengths of greater than two, so there's no
contradiction with a result for node degree.  There is clearly freedom
in defining the details too. The formula is based on a spherical
geometry from the origin, with radius $r$ equal to the graph hop
count.  In particular, when using the number of hops from a starting
point as the only measurable form of distance, there is some freedom
in imagining the filling geometry (see figure \ref{cubes}), because
the result is effectively an open ball without a boundary of corners
or faces, so there's a discrepancy in calculating volume from nodes or
links.  For a Cartesian `orthogonal' geometry, we would tend to count
volumes in terms of a cube of side-length $2r$; now, the paths of length $R$
may undergo right angle turns---they can't be assumed purely radial, which
complicates the notion of distance by hop alone (this is one reason why
many researchers immediately look to embed graphs in a Euclidean container).
To complete what we think of as a cubic volume, we may have to go through
additional hops to include corners, so a simple radial path length
will progressively underestimate dimension in a geometry that
undergoes right angle turns at the end, as a function of the distance.
A counting estimate actually lies closer to a count from one hop
length greater (at least for small hop counts and dimensions), so

There are two possible methods for using (\ref{dimdef}). 
\begin{enumerate}
\item We can
measure the dimension as a function of distance (number of hops
traversed along all paths of fixed length from the source) and sum the
nodes enclosed by this region. This gives a deterministic answer for the dimension as
a function of scale at that point, due to the Long Range Order of the lattice.
\item We can select a random sample of points and perform the same method, but
now average over the results to give a statistical estimate of a wider area.
\end{enumerate}

\section{Scale dependence of statistical measures}
The scale dependence of the results is of interest here. In the middle
of an $n$-cube, far away from edge effects, we would expect the
dimension to be quite stable. If we think of the volume of an $3$-cube as $L^3$
and $L=2$, then the dimension would be $\log(8)/\log(2) = 3$, which makes sense.
But this is not how we measure volume from a graph (see figure \ref{cubes}). The corner
nodes are not part of the volume found at one hop in all directions (giving a side length of 2).
Only the open core is counted. There is no way to speculate about completing some kind of idealized
geometry to get this nice simple number. Instead, we have to live with
empirical and fractional approximations from inside the graph.

Closer to an edge we might expect the dimensionality of a volume $d$ to be reduced by
the effect of edges. This is the case, but it never reaches the
dimension of the face ($d-1$).  More interesting, we find that the dimension
reaches a maximum value, due to the efficiency of dividing a virtual
volume into balls, which we may then fall off again with distance (see
figures \ref{hyper2} and \ref{hyper3}) to reach a stable value as long
as we don't get close to edges.

A Euclidean spatial lattice has a singular degree distribution (a delta
function at some fixed number of nearest neighbours). However, 
such networks are special cases, typically due to some Long Range Order; 
few networks that represent processes have such uniform crystalline bindings.
Internet is known to have a power law distribution of node degrees,
which means that error bars are only convenient scale markers, not
Gaussian measures, and they will be large.

In inhomogeneous graphs, the non-simply connected topologies may lead to dimension
being different on short and long scales, due to the inherent scale of
Long Range Order (molecules or cells are not uniform on a short scale, but may
form materials that are uniform on a large scale). This presents a dilemma: we restrict the
size of measurement to capture local variations, while trying to avoid
data that are merely incomplete. The only solution is to probe as a
function of length and use any emergent scales to identify a natural
granularity in space. Within grains of a certain size, there can be
``hidden extra dimensions'' or the opposite ``holes through the
dimensionality'' that affect the result on a smaller scale. This is it
the graphical analogue of ``hidden dimensions'' known from speculative
physics models, e.g. Kaluza-Klein theory.

For anyone who believes the true nature of spacetime to be some kind
of metric space, this inconsistency of dimensionality might feel
weird, but this is what one has to expect of any discrete process
structure. The infinitesimal limit (to which we've become
unquestioningly accustomed) is the only case in which dimensionality can be abstracted
to a constant. As we move towards theories in which the primacy of processes over 
things and spaces begins to dominate, this strangeness will become simply natural.

We can calibrate the expression defining scale-dependent graph dimension using $n$-dimensional hypercubes, which are easy to
generate as lattice graphs $\Gamma(\nu,\lambda)$. 
For a cube of discrete side length $L$ (i.e. $L$ links joining
$L+1$ nodes along an edge), we can calculate the numbers of links $\lambda$ and nodes $\nu$:
\beq
\nu(L) &=& (L+1)^n\\
\lambda(L) &=& nL(L+1)^{n-1}.
\eeq
Using these measures, we can check the correctness of the graph and the links in data,
as well as perform Taylor expansions to estimate the discrepancy between an idealized cube
and an effective spanning tree volume.
A graph database (here we used ArangoDB) makes this task quite easy,
with machine validation for large graphs.
\begin{table}
\begin{center}
\begin{tabular}{|c|c|c|}
\hline
$n_\text{Euc}$ & $\langle D_n\rangle \pm \sigma(D_n)$ & $\max(D_{n})$\\
\hline
2 & $2.21 \pm 0.06$ & 2.3\\
3 & $3.03 \pm 0.07$ & 3.1\\
4 & $3.76 \pm 0.16$ & 4.3\\
5 & $4.46 \pm 0.12$ & 5.2\\
ITDK-dev & $6.12 \pm 0.00$ & 8.7\\
ITDK AS & $5.16 \pm 0.00$ & 7.70\\
\hline
\end{tabular}
\caption{\small Estimated dimension after 20 hops in a hypercube of side length 20 from randomized starting locations. After 20 hops, the dimension continues to fall off much more slowly within the margins of error, except for the AS map, where the dimension continues to fall off as a function of distance because
no new points get added after the confinement length of 7-8 hops, thus the dimension is diluted and we should truncate at 8 hops.}
\end{center}
\end{table}

When dealing with expressions involving powers, it's hard to find an expression
that doesn't require one to know the power in order to find the power, so we resort
to linearizing the expressions with logs and expansions.
Ultimately, we can choose simply to define a heuristic statistical dimension $D$,
based on hop lengths from an origin, by:
\beq
D \equiv \frac{\log (N_\text{nodes})}{\log (N_\text{hops}-\2+1/N_\text{hops})},
\eeq
where the additional corrective terms in the denominator are the first
terms in a Taylor expansion for large $L$, and offer an approximate
calibration to cubic lattices up to dimension 5. Since our intuition about dimension
is limited to these low numbers, this seems like a fair compromise.

\section{Semantic channels}

Figure \ref{itdk_model} shows the different channels for which data
were collected in the ITDK probes.  Device nodes are the basic meeting
point for information channels. They form the interior structure of AS
nodes, for example, so they form an entirely different metric and
semantic scale of the functional behaviour.  There is an independent
set of IPv4 and IPv6 addresses (the latter currently empty in the
data).  Finally, we have Autonomous System Numbers (ASN) and partial
information about geo-spatial coordinates for the city a node is close to---a
very coarse location on the Earth's surface.  We examine only Devices and
ASes here.  The dimensionality of these different channels need not be
the same.

Although the probe process trajectories are {\em directed}
(directional) in nature, the ITDK data treat connections as symmetric
or undirected, in keeping with the usual interpretation of a
background space.  This is a weakness in the method for the ITDK,
which assumes that reflection paths are the same as outgoing paths.
This is not true in general, but there is currently no way to
determine this using available technology.  Locations are the average
remnants of processes that could move in any direction (the process
memory, in a computational sense). Formally, we use NEAR type links of
the Semantic Spacetime model as correlations with respect to ensembles
of probe trajectories.

In the case of the AS channel, connections are much more sparse and we
need to collect an order of magnitude more samples in order to get an
estimation of the error. This is intuitively in keeping with the fact
that, if we call Device nodes semantic scale $S^{n}$ then AS has
logarithmic scale $S^{n+1}$.  More interesting, perhaps, the AS graph
seems to exhibit {\em confinement}, i.e.  there is a maximum path
length: paths don't propagate beyond 7 or 8 hops.  The reason for this
is unclear, and we can't rule out the idea that it is an artifact of
the data collection limitations. The Autonomous Systems (AS) are
private organizations, and may not allow probing of their interior
structures. The interiors may also be connected through `Unknown'
nodes, which are tunnels or wormholes through the Internet,
unobservable jumps of often indeterminate length. On the other hand,
one would expect the maximum number of hops between all reachable
locations to be finite and the range is shorter on the scale of ASes
than for Devices.  Evidence of maximum ranges above 20 hops were also
seen in the Devices channel, but the computational expense or going
beyond this prohibited further study. It's at least suggestive that the finite horizon
is a real effect rather than an artifact of the data.

\begin{figure}[ht]
\begin{center}
\includegraphics[width=9cm]{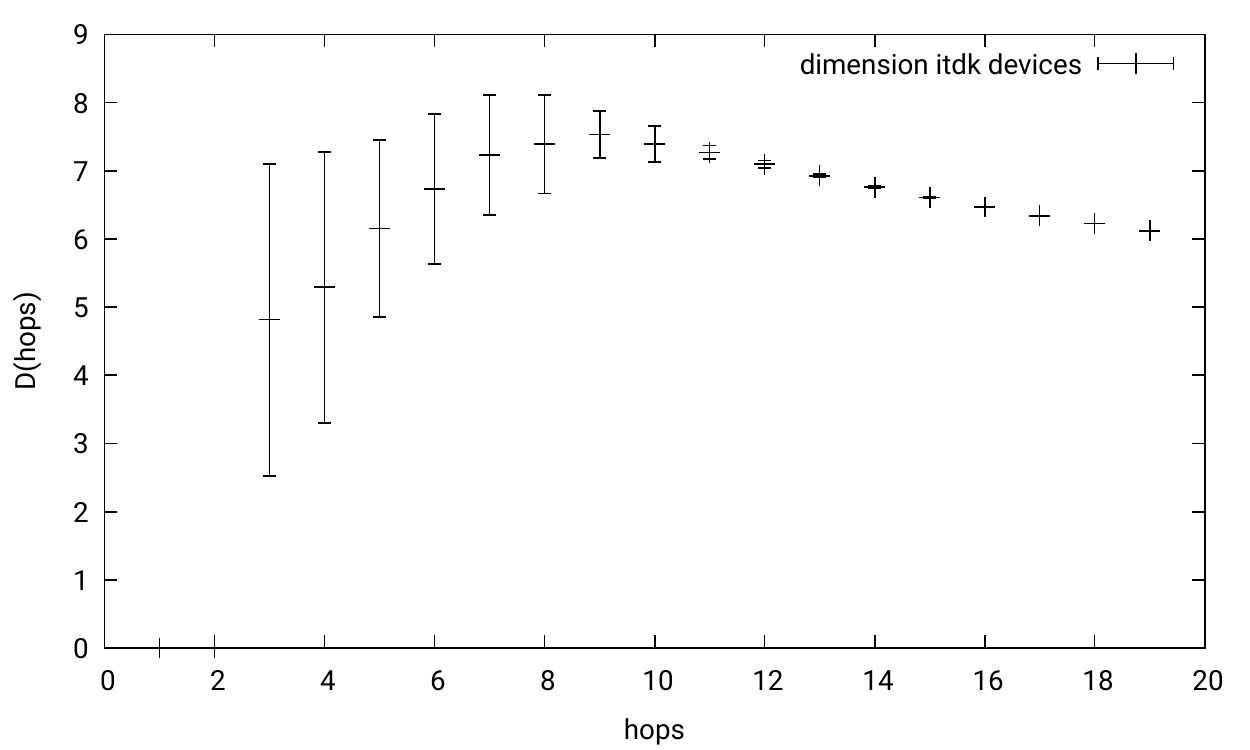}
\caption{\small Dimension of the device and IPv4 network as a function of hops from a random origin, based on ITDK data. The calibrated dimension is $6.12 \pm 0.00$\label{itdk_dim}}
\end{center}
\end{figure}

\begin{figure}[ht]
\begin{center}
\includegraphics[width=9cm]{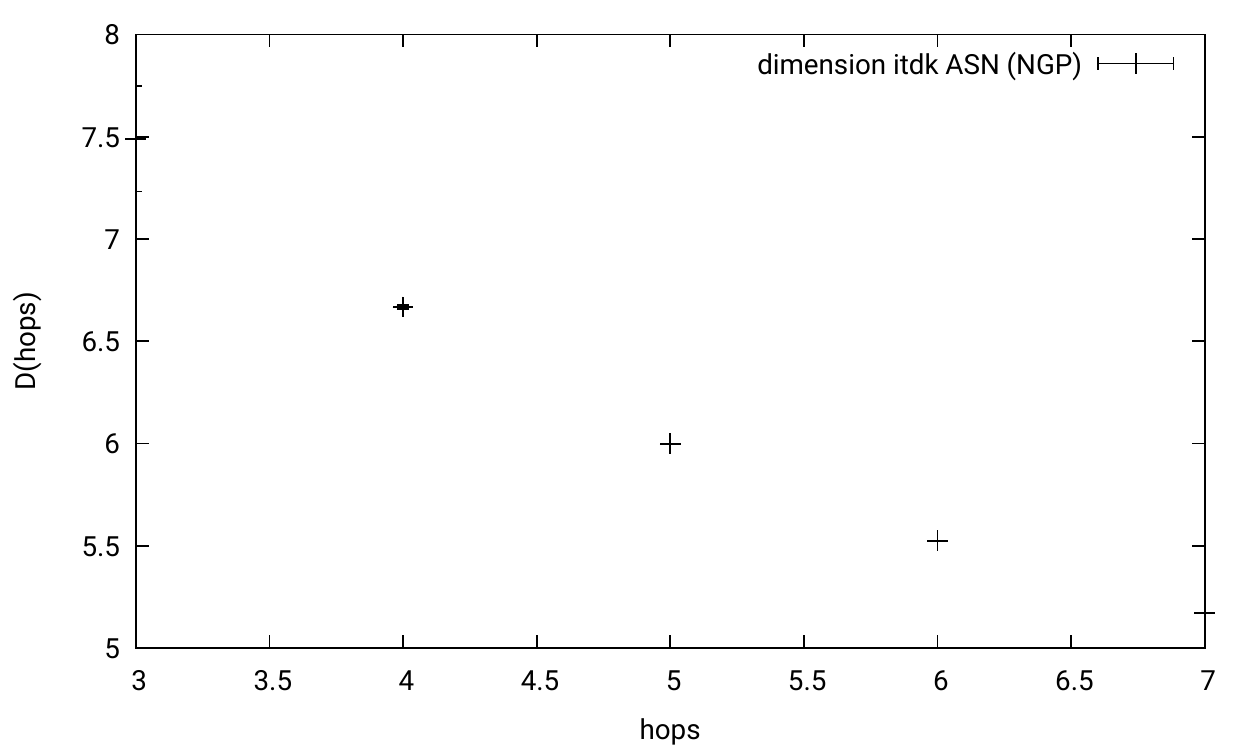}
\caption{\small Dimension of the AS (BGP) scale as a function of hops from a random origin, based on ITDK data. The calibrated dimension is $5.17 \pm 0.00$\label{itdk_AS}}
\end{center}
\end{figure}

The AS supernode graph has a fundamentally different character to the
device graph.  Cluster radii of less than 8 hops characterize the
observable data. At 8 hops, the effective dimension appear to be about
5. This continues to fall with hop length, because no new points are
expressed into the expanding volume of the search, thus it falls off
exponentially without a clear limit. That is not the say that there
are no new AS bubbles beyond this distance, only that they are
disconnected by paths that are not out of scope.  Some of these could
be connected by anonymous tunnels (wormholes or hyperlinks) which have
not been considered here because their passage through ASes can't be
established.

\section{Summary}

In a dynamical system, we have to be clear about the distinction
between infrastucture carriers and the virtual processes that multiply
on top of it, over different semantic channels. We can only ever
characterize processes using other processes, so at one level
everything is virtual. This relativity trap may require us to modify
certain prejudices about the nature of reality from observation alone.

It's possible to calibrate and measure the dimension of the Internet
across two of its process channels: device nodes and ASes.  The
spacelike hypersections of the Internet's {\em routing device channel}
have a short range dimension rising from 1.65 to an intermediate cell
size of 8.7.  Over distances at around 20 hops, the effective average
dimension settles to $D(20) = 6.12 \pm 0.00$. The scale dependence is evidence of the
`foamy' honeycomb structure of the graph.  For the AS channel,
there is a similar pattern and evidence of containment at a maximum length of paths is no more than 7 or
8 hops.

Last mile `edge' connections of devices are conspicuously absent from the
data in the map. This includes subnets where broadcast domains effectively group endpoints into
coarse grains of order $10^2$ devices. This number of direct edge to
edge tunnelling is growing all the time, and will likely be impacted
in reality by telecommunication 5G services and beyond. 

We can thus compute the dimension of the Internet as an abstract
spacetime. The discreteness of the Internet scale favour a higher
number of dimensions than a continuum embedding space would, since it
is unconstrained by continuity or analyticity.  At around six
dimensions per slice, and higher dimensionality on a small scale, we
see that it is quite `foamy' more like a honeycomb than a uniform
Euclidean space. On a practical level, this means that for every
possible route between two points, there is an average redundancy of
about 6 alternatives. On a policy level, this indicates a measure of
the robustness and even resilience of the Internet to outages.

Due to the absence of timelike process information in the ITDK sets,
we can't comment on the long timescale dimension for slow changes.
It's not clear how to measure this timelike evolution of the Internet
as a whole, in either a Newtonian or Minkowskian sense. The proper
time of a process is intrinsically local and short range. The global
time assumption is a convenient mathematical approximation for an
embedding space, but it isn't really physical within the interior of a system.
Locally, individual observations express
time inseparably from space, and may have any dimensionality (in
principle) between 1 and some natural number. The default traceroute
probe is to send three probes in parallel to initiate three parallel
paths, subject to underlying routing behaviour; but over long
distances observations those probe paths grow only with a lower number
of dimensions over the expanding cone of reachable states.  The
computational expense of exploring path lengths of longer than 20
nodes is very high, even with a sparse graph, since path number grows
exponentially with length.

The study here is limited by the maps we can construct of Internet
processes.  This, in turn, is limited by politics and security policy, as well
as by technology.  The Internet wasn't explicitly designed to be
observed (much like the universe as a whole) so we have to make do
with what we have. Recognizing certain dynamical similarities between
virtual network processes and quantum systems, we could imagine a
future methodology based on interferometry for revealing more of the
dynamical effects; however, this remains in the realm of speculation
for now. 

I'm grateful to kc claffy and Bradley Huffaker of CAIDA for helpful discussions
and for the use of their resources.

\bibliographystyle{unsrt}
\bibliography{spacetime}

\end{document}